\documentclass{llncs}

\usepackage{enumerate}

\usepackage{amsmath}
\usepackage{amsfonts}

\spnewtheorem*{algorithm}{Algorithm}{\bfseries}{}

\newcommand{\unif}{\stackrel{?}{=}}
\newcommand{\xor}{\oplus}
\newcommand{\std}{\text{std}}
\newcommand{\ACUN}{\text{\normalfont{}ACUN}}
\newcommand{\crypt}[2]{\{#1\}_{#2}}
\newcommand{\inv}[1]{#1^{-1}}
\newcommand{\pair}[2]{\langle{#1},{#2}\rangle}

\begin{document}

\pagestyle{headings}

\mainmatter              

\title{Implementing a Unification Algorithm for Protocol Analysis with XOR}
\author{Max~Tuengerthal\inst{1} \and Ralf~K\"usters\inst{1} \and Mathieu~Turuani\inst{2}}
\institute{Christian-Albrechts-Universit\"at zu Kiel, Germany\\ \email{mtu@informatik.uni-kiel.de,kuesters@ti.informatik.uni-kiel.de}
 \and Loria-INRIA,
  Vandoeuvre-l\`es-Nancy, France\\
\email{mathieu.turuani@loria.fr}}

\maketitle

\section{Introduction}\label{sec:Introduction}

Many methods and tools for the fully automatic analysis of security protocols
are based on a technique called constraint solving (see, e.g.,
\cite{MillenShmatikov-CCS-2001,ChevalierVigneron-ASE-2001}), which as a central
component involves a unification algorithm. The first methods and tools for the
analysis of security protocols assumed the message space to be a free term
algebra. However, this is a too idealized assumption in case the protocols
employ operators involving algebraic properties, such as the exclusive or
(XOR), an operator frequently used in security protocols.  In
\cite{ChevalierKuestersRusinowitchTuruani-LICS-2003,ComonShmatikov-LICS-2003}
it was shown that the security, more precisely secrecy and authentication, of
protocols is still decidable w.r.t.~a bounded number of sessions, even
NP-complete \cite{ChevalierKuestersRusinowitchTuruani-LICS-2003}, when taking
algebraic properties of XOR into account.  However, these results do not yield
practical algorithms. A first algorithm based on constraint solving and
tailored towards efficient implementation was proposed by Chevalier
\cite{Chevalier-UNIF-2004}.  However, a prerequisite for this algorithm to be
of practical use is a unification algorithm for a combination of the equational
theory $E_\ACUN$ (modeling algebraic properties of XOR) and an equational
theory $E_{\std}$ modeling public/private keys which works well in practice.
The goal of the present work is to provide such an algorithm.

A unification algorithm for $E=E_{\std}\cup E_{\ACUN}$ can easily be obtained
by the general combination method proposed by Baader and Schulz
\cite{BaaderSchulz-JSC-1996}, since unification algorithms for $E_{\std}$ and
$E_{\ACUN}$ exist. However, this unification algorithm would be highly
non-deterministic and therefore not directly suitable for practical use.
Several optimizations have been proposed. First, Baader and Schulz
\cite{BaaderSchulz-JSC-1996} already suggested simple optimizations. More
sophisticated optimizations, called iterative and deductive method, were
presented by Kepser and Richts \cite{KepserRichts-FCS-1999}, who exploit
concrete properties of the theories, like collapse-freeness, to limit the
non-determinism. Another combination method, along with optimizations, was proposed by
Boudet~\cite{Boudet-JSC-1993}. However, the settings in all of these works are
still quite general and their optimizations do not suffice for our purposes.

In this paper, we propose a unification algorithm for the theory $E$ which
combines unification algorithms for $E_{\std}$ and $E_{\ACUN}$ but compared to
the more general combination methods mentioned above uses specific properties
of the equational theories for further optimizations. Our optimizations
drastically reduce the number of non-deterministic choices, in particular
those for variable identification and linear orderings. This is important for
reducing both the runtime of the unification algorithm and the number of
unifiers in the complete set of unifiers. We emphasize that obtaining a
``small'' set of unifiers is essential for the efficiency of the constraint
solving procedure within which the unification algorithm is used.

\medskip

\noindent \emph{Outline of the Paper.} In the following section, we briefly
recall the combination algorithm by Baader and Schulz along with the
optimizations proposed by Kepser and Richts. In Section~\ref{sec:Our Optimized
  Algorithm}, our unification algorithm is introduced, with experimental
results presented in Section~\ref{sec:Experimental Results}. We conclude in
Section~\ref{sec:Conclusion}. Further details can be found in a technical
report \cite{Tuengerthal-IFI-TR-0609-2006}.

\section{The General Combination Algorithm}\label{sec:The General Combination Algorithm}

In this section, we briefly describe the general combination method of Baader
and Schulz \cite{BaaderSchulz-JSC-1996} and optimizations introduced by Kepser
and Richts \cite{KepserRichts-FCS-1999} as our algorithm is based on
\cite{BaaderSchulz-JSC-1996} and some optimizations are motivated by
\cite{KepserRichts-FCS-1999}.

Given disjoint equational theories $E_1$ and $E_2$ and stand-alone unification
algorithms $A_1$ and $A_2$ for $E_1$ and $E_2$, respectively, which work with
linear constant restrictions (see below) the combination method of Baader and
Schulz combines $A_1$ and $A_2$ to obtain a unification algorithm for the
joined theory $E = E_1 \cup E_2$.  More precisely, given an elementary
$E$-unification problem $\Gamma$, the combination method works as follows:
\begin{enumerate}[1.]
\item\label{variable abstraction old} \textbf{Purification and splitting}.
  Obtain the sub-problems $\Gamma_{1,x}$, with $x\in \{1,2\}$, by purifying
  terms and splitting equations for each theory $E_x$. (Non-pure terms or
  equations are those containing symbols of different theories.)
  
\item\label{variable identification} \textbf{Variable identification}. Choose a
  partition (i.e., equivalence classes) on variables for each $\Gamma_{1,x}$,
  $x\in \{1,2\}$. Let $\Gamma_{2,x}$ be the sub-problem obtained from
  $\Gamma_{1,x}$ by replacing each variable by a representative of its class.
  
\item\label{choose theory indices} \textbf{Choose theory indices}.  For each
  variable $v$ in $V$ choose a theory index $Ind(v)\in \{1,2\}$ where $V$ is
  the set of variables occurring in both $\Gamma_{2,1}$ and $\Gamma_{2,2}$. If
  in $\Gamma_{2,1}$ a variable has theory index 2 it is considered a constant
  in $\Gamma_{2,1}$; analogously for $\Gamma_{2,2}$.
  
\item\label{choose linear ordering} \textbf{Choose linear ordering}. Choose a
  linear ordering $<$ on $V$. (Together with 3., the linear ordering $<$ induces
  what Baader and Schulz call a \emph{linear constant restriction}.)
  
\item\label{solve system} \textbf{Solve systems}.  For each theory $E_x$, the
  algorithm $A_x$ is applied to $\Gamma_{2,x}$ and $<$ to produce a complete
  set $C_x$ of unifiers respecting $<$, where a unifier $\sigma$ respects $<$
  if $x<y$ implies that $y$ does not occur in $x\sigma$ for every $x,y\in V$.
  
\item\label{combine unifiers} \textbf{Combine unifiers}.  If $C_1$ or $C_2$ are
  not empty, combine the unifiers of $C_1$ with those of $C_2$ to obtain a set
  of $E$-unifiers of $\Gamma$. Go back to 2.~to try other choices (in order to
  obtain further unifiers).
\end{enumerate}

\begin{theorem} \cite{BaaderSchulz-JSC-1996}\label{thm correctness of general method}
The set of $E$-unifiers produced by the combination method above form a complete set of $E$-unifiers of the $E$-unification problem $\Gamma$.
\end{theorem}
The major disadvantages of the general combination method are its high degree
of non-determinism and the non-detection of failures before the last step. This
results in poor runtime behavior and sets of unifiers that are far from
minimal.

The main idea of the optimizations of Kepser and Richts
\cite{KepserRichts-FCS-1999} are to first make all non-deterministic decisions
for one component in order to detect failures as soon as possible (iterative
method) and to use constraints obtained by solving one component for reducing
the number of remaining non-deterministic choices (deductive method).

\section{Our Optimized Algorithm}\label{sec:Our Optimized Algorithm}

We now present our unification algorithm for the equational theory $E = E_\std
\cup E_\ACUN$ where $E_\std = \{ x \approx \inv{(\inv{x})}\}$ with $\inv{\cdot}$
modeling a mapping between public and private keys and $E_\ACUN = \{ x \xor (y
\xor z) \approx (x \xor y) \xor z, x \xor y \approx y \xor x, x \xor 0 \approx
x, x \xor x \approx 0\}$ for modeling the XOR operator. The theory $E_{\std}$
is associated with a signature containing finitely many free symbols of
arbitrary arity, including constants or binary symbols for pairing
$\pair{\cdot}{\cdot}$ and encryption $\crypt{\cdot}{\cdot}$. The signature
associated with $E_\ACUN$ is $\{\xor, 0\}$. We note that both $E_\std$ and
$E_\ACUN$ are unitary for elementary unification and efficient unification
algorithms exist for both theories. However, it is not hard to see that $E$ is
not unitary; by Theorem~\ref{thm correctness of general method} $E$ is
finitary.  Unification for $E$ can easily be shown to be NP-complete using
results in \cite{GuoNarendranWolfram-CADE-1996}.

In what follows, we summarize the main optimizations of our algorithm compared
to those discussed in the previous section, along with brief justifications of
their correctness. Our optimizations employ specific properties of the
equational theories under consideration and they reduce both the runtime and
the size of complete unification sets.

\medskip\noindent\textbf{Simplified iterative and deductive method.}  Similar
to Kepser and Richts, we employ the idea of the iterative and deductive method
but apply it only once to $E_\std$. That is, we first solve the
$E_\std$-unification problem without any constraints. If this fails, the
original problem is unsolvable. Otherwise, we obtain an mgu $\sigma_\std$ used
in subsequent steps to reduce the number of non-deterministic choices. Since
typically the $E_\ACUN$-unification problem will not yield further constraints,
we postpone solving this unification problem to a later point.

\medskip\noindent\textbf{Hierarchy of variable identifications.}  A major new
optimization in our algorithm is that we do not have to iterate over all
possible variable identifications. If unification for both $\Gamma_\std$ and
$\Gamma_\ACUN$ succeeds for some variable identifications $p$ and $p'$ where
$p$ is more general than $p'$, then the combined unifier for $p$ is more
general than the one for $p'$.  This can be shown using the following property
of $E_{\ACUN}$:
\begin{lemma}\label{lem:propertyACUN}
  Every mgu of a $E_\ACUN$-unification problem with linear constant restriction
  is also an mgu of this unification problem without restrictions.
\end{lemma}
The above property on variable identifications allows us to traverse the tree
of variable identifications in a breadth-first manner and skip all less general
variable identifications once we succeed in solving the problem for a more
general one.

\medskip\noindent\textbf{Reduce number of choices of indices.}  Most theory
indices can be determined from $\sigma_\std$. If a variable is instantiated by
a term with a collapse-free top-symbol, then this variable has to be a constant
in $\Gamma_\ACUN$.  On the other hand, if $x$ is not instantiated by
$\sigma_\std$ and if there exists no variable $y$ with $y\sigma_\std =
\inv{x}$, then it does not matter whether $x$ is treated as a constant in
$\Gamma_\std$ or not.  In fact, a non-deterministic choice of theory indices
must only be made for variables $x$ and $y$ such that $x\sigma_\std = \inv{y}$
and $y\sigma_\std=y$.  Of course, not both can be constants in $\Gamma_\std$,
so it suffice to choose one of them.

\medskip\noindent\textbf{Reduce number of choices of linear orderings.}  Instead
of choosing an arbitrary linear ordering on $V$ (see Section~\ref{sec:The
  General Combination Algorithm}), we first deduce (deterministically) a
partial ordering $<_{po}$ from $\sigma_\std$ such that $x <_{po} y$ iff $y$
occurs in $x\sigma_\std$.  Now, the important observation is that by
Lemma~\ref{lem:propertyACUN} once we have found a solution of the
$E_{\ACUN}$-unification problem w.r.t.~a linear ordering $<$ which extends
$<_{po}$, we do not need to try other linear orderings.

\begin{theorem}
  The algorithm described above returns a complete set of $E$-unifiers for a
  given $E$-unification problem.
\end{theorem}
We note that the optimizations explained above are fairly independent of the
theory $E_\std$. Hence, $E_\std$ can easily be replaced by other theories.

\section{Experimental Results}\label{sec:Experimental Results}

\begin{table}[t]%
\caption{Runtimes and sizes of complete sets of unifiers: ``size'' denotes the size of the returned complete set of
  unifiers; ``vi opt'' stands for ``variable identification optimization'';
  $x,y,z,u,x_i$ are variables and $a,b,c,d,e$ are constants. Runtime 
  tests obtained on a 1.5\,GHz Intel Pentium M processor.}
\begin{center}
\begin{tabular}{r@{\quad}l@{\quad}r@{\quad}r@{\quad}r@{\quad}r}
\hline
\rule{0pt}{9pt} & & \multicolumn{2}{c@{\quad}}{with vi opt} & \multicolumn{2}{c}{without vi opt}\\
\multicolumn{1}{c@{\quad}}{no} & \multicolumn{1}{c@{\quad}}{unification problem} & \multicolumn{1}{c@{\quad}}{\parbox[c]{30.5pt}{\rule{0pt}{10pt}\centering time (msecs)}} & \multicolumn{1}{c@{\quad}}{size}  & \multicolumn{1}{c@{\quad}}{\parbox[c]{30.5pt}{\rule{0pt}{10pt}\centering time (msecs)}} & \multicolumn{1}{c}{size}\\[9pt]
\hline
1&\parbox{145pt}{$\pair{x}{\pair{\crypt{x \xor y}{a}}{\crypt{\crypt{x \xor y}{a} \xor z}{a}}} \unif_E$\newline \hspace*{1em} $\langle\crypt{b \xor c}{a},\langle\crypt{\crypt{b \xor c}{a} \xor d}{a},$\newline \hspace*{1em} $\crypt{\crypt{\crypt{b \xor c}{a} \xor d}{a} \xor e}{a}\rangle\rangle$} & 3.3 & 1 & $>$ 30\,min & \\ 
2&$z \unif_E \crypt{\pair{x}{\pair{y}{x \xor y}}}{\inv{(z \xor u)}}$ & 0.1 & 1 & 3.3 & 15\\
3&$z \unif_E \crypt{\pair{x}{\pair{y}{x \xor y}}}{\inv{(z \xor a)}}$ & 9.1 & 0 & 9.1 & 0\\
4&$0 \unif_E \pair{x_1}{y_1} \xor \dots \xor \pair{x_9}{y_9}$ & 9.6\,s & 0 & 9.6\,s & 0\\
5&$0 \unif_E \pair{x_1}{y_1} \xor \dots \xor \pair{x_{10}}{y_{10}}$ & 239.7\,s & 945 & 70.7\,s & 6556\\[2pt]
\hline
\end{tabular}
\end{center}
\label{tab:some-exmp}
\end{table}

Table~\ref{tab:some-exmp} summarizes some of our experimental results (see
\cite{Tuengerthal-IFI-TR-0609-2006} for more). It contains runtimes and sizes
of complete sets of unifiers both with the optimization for variable
identification turned on and off. (The other optimizations are harder to turn
on and off in our implementation, which is why these optimization are always
turned on.) These results show that our unification algorithm runs efficiently
on many benchmarks and that our optimizations indeed reduce both runtime and
size of complete sets unifiers. In fact, the optimized version of our algorithm
always returned minimal sets of unifiers. (However, we have no proof that this
is always the case.)

Problem 1 in Table~\ref{tab:some-exmp} is a unification problem that occurs in
the analysis of the recursive authentication protocol
\cite{RyanSchneider-IPL-1998}. Interestingly, while our algorithm quickly
returns an mgu, the version of the algorithm with the optimization for variable
optimization turned off does not come back with a solution within 30 minutes.
The two versions of the algorithm also perform very differently on problem 2.
There is no difference in problem 3 since this problem is not unifiable, and
hence, the algorithm has to try all possible variable identifications. Problems
4 and 5 are only of theoretical interest, they typically do not occur in
applications but illustrate the limitations of optimizations. Note that in
problems of this form the size of a minimal complete set of unifiers may be
exponential in the size of the problems.

\section{Conclusion}\label{sec:Conclusion}

Motivated by the analysis of security protocols, we have presented a
unification algorithm for an equational theory including ACUN. Our algorithm
contains several optimizations which make use of the specific properties of the
equational theories at hand and performs well on practical examples, both in
terms of its runtime and the size of the complete set of unifiers returned. As
such, our algorithm is well-suited as a subprocedure in constraint solving
algorithms for security protocol analysis with XOR.

One future direction is to incorporate other operators and their algebraic
properties into our algorithm, including important operators such as
Diffie-Hellman Exponentiation and RSA encryption. In
\cite{ChevalierKuestersRusinowitchTuruani-FSTTCS-2003,ChevalierKuestersRusinowitchTuruani-ENTCS-2005},
it was shown that fully automatic analysis of security protocols is also
possible in presence of such operators.


\end{document}